\documentclass[12pt]{article}

\usepackage{amssymb,graphicx,url,colordvi}
\usepackage[affil-it]{authblk}
\usepackage[super]{natbib}
\usepackage{graphicx,color,amsmath}
\usepackage{soul,xcolor}

\newcommand\arcsec{\mbox{$^{\prime\prime}$}}

\newcommand\apjl{The Astrophysical Journal Letters}
\newcommand\aap{Astronomy \& Astrophysics}
\newcommand\apj{The Astrophysical Journal}

\newcommand\ao{Applied Optics}

\newcommand\solphys{Solar Physics}
\newcommand\mnras{Monthly Notices of the Royal Astronomical Society}
\newcommand\nat{Nature}
\newcommand\ssr{Space Science Reviews}
\newcommand\araa{Annual Review of Astronomy and Astrophysics}
\newcommand\procspie{Proceedings of the SPIE}

\begin{document}
\title{\textbf{Unprecedented Fine Structure of a Solar Flare Revealed by the 1.6~m New Solar Telescope}}

\author[1,2]{Ju Jing\thanks{Correspondence and requests for materials should be addressed to J.J. (email: ju.jing@njit.edu)}}
\author[1,2]{Yan Xu}
\author[1,2]{Wenda Cao}
\author[1,2]{Chang Liu}
\author[1,2]{Dale Gary}
\author[1,2]{Haimin Wang}

\affil[1]{Center For Solar-Terrestrial Research, New Jersey Institute of Technology, University Heights, Newark, NJ 07102-1982, USA}
\affil[2]{Big Bear Solar Observatory, New Jersey Institute of Technology, 40386 North Shore Lane, Big Bear City, CA 92314-9672, USA}

\renewcommand\Authands{ and }

\maketitle

\setstcolor{red}

\begin{abstract}
\normalsize
Solar flares signify the sudden release of magnetic energy and are sources of so called space weather. The fine structures (below 500 km) of flares are rarely observed and are accessible to only a few instruments world-wide. Here we present observation of a solar flare using exceptionally high resolution images from the 1.6~m New Solar Telescope (NST) equipped with high order adaptive optics at Big Bear Solar Observatory (BBSO). The observation reveals the process of the flare in unprecedented detail, including the flare ribbon propagating across the sunspots, coronal rain (made of condensing plasma) streaming down along the post-flare loops, and the chromosphere's response to the impact of coronal rain, showing fine-scale brightenings at the footpoints of the falling plasma. Taking advantage of the resolving power of the NST, we measure the cross-sectional widths of flare ribbons, post-flare loops and footpoint brighenings, which generally lie in the range of 80-200 km, well below the resolution of most current instruments used for flare studies. Confining the scale of such fine structure provides an essential piece of information in modeling the energy transport mechanism of flares, which is an important issue in solar and plasma physics.

\end{abstract}

\section*{Introduction}
Ever since a solar flare was first detected by Carrington\cite{Carrington1859} and Hodgson\cite{Hodgson1859} in 1859, this spectacular phenomenon of solar activity has been a subject of intense research and has served as a natural laboratory for understanding the physical processes of transient energy release throughout the universe. From the early solar telescopes in Carrington and Hodgson's day to those of today, great advances in observing facilities and overcoming the turbulence of Earth's atmosphere have continued to improve spatial resolution and provided much greater resolving power to study flares in much more detail.

Based on such observations, solar physicists have established theories, models and simulations to explain the physical processes in solar flares and their stellar analogues. In the standard flare model\cite{Carmichael1964, Sturrock1966, Hirayama1974, Kopp1976}, the primary energy release and subsequent particle acceleration occur in the corona as a result of magnetic reconnection\cite{Holman2011}. The precipitating nonthermal particles from the reconnection site, and thermal conduction from the reconnected loops, are two agents for transporting the energy released in the corona to the lower and denser chromosphere where most of the energy is finally converted into heat and radiated away as thermal emission\cite{Benz2010}. However, many of the physical details of this standard model and subsequent processes are presently not well understood, or are observationally unconstrained, due to the limit of spatial resolution. For example, the flux of nonthermal electron beams is important, yet is poorly constrained in the framework of the model for understanding the energy transport mechanism during flares. This flux not only is sensitive to the energy distribution of precipitating electrons but also depends critically on an accurate measure of the emission area, thus necessitating a higher resolution to resolve the fundamental spatial scale of flare emission. In addition, many numerical simulations and models of magnetic reconnection suggest the existence of unresolved fine-scale coronal loop components to reproduce observed warm coronal loops\cite{Gomez1993, Reale2000, Beveridge2003, Warren2003, Klimchuk2006, Klimchuk2008, Brooks2012}, which is awaiting higher spatial resolution observation to verify it.

The new generation of large, i.e., meter-class aperture, ground-based solar telescopes is now entering the scene to make fundamental discoveries. Many unresolved scientific issues including the above mentioned could benefit greatly from these powerful tools. For instance, taking advantage of the high resolution observation (with a pixel size of 0\arcsec.0597 corresponding to $\sim$40 km) of CRisp Imaging Spectro-Polarimeter (CRISP)\cite{Scharmer2008} at the Swedish 1~m Solar Telescope (SST)\cite{Scharmer2003}, Antolin \& Rouppe van der Voort (2012)\cite{Antolin2012} and Scullion et al. (2014)\cite{Scullion2014} detected the coronal loop substructures which are on $\sim$100 km spatial scales. The 1.6~m New Solar Telescope (NST)\cite{Goode2012} at Big Bear Solar Observatory (BBSO) is another large telescope known for its leading-edge observational capabilities. With the adaptive optics system and the speckle image reconstruction processing technique\cite{Woger2007}, the Visible Imaging Spectrometer (VIS)\cite{Cao2010} at NST can achieve a pixel size as small as $\sim$0\arcsec.03 (corresponding to $\sim$ 20 km), unparalleled by any previous observation. An earlier NST/VIS H$\alpha$ observation revealed a fine structure of flare ribbons in a sunspot that consists of a string of bright knots as small as $\sim$100 km\cite{Sharykin2014}. To our knowledge, this is the finest-resolution flare ribbon reported to date. However, due to limited duty cycle in good seeing conditions, such a diffraction-limited flare observation is extremely rare.

In this paper we present a high spatial resolution observation of an M-class flare using the NST/VIS chromospheric H$\alpha$ images. In addition to the long-enduring flaring process, the observation also reveals details of coronal rain (made of condensing plasma during the post-flare cooling phase)\cite{Muller2003, Moschou2015} falling down along the post-flare loops, and the chromosphere's response to the impact of coronal rain, showing fine-scale brightenings at the footpoints of the falling plasma. Using images of multiple wavelength in the H$\alpha$ spectral line, we measure the cross-sectional widths of the flare ribbons, the post-flare loops and the brightenings at the footpoints, all of which are generally on sub-acrsecond scales ($<$ 200 km). Our observation provides a novel information on the spatial scale of the energy transport and heating mechanism of solar flares.

\section*{Observations and Results}
The flare that we discuss in this paper appeared in NOAA active region (AR) 12371 on 2015 June 22, with a peak GOES soft X-ray (SXR) flux at 18:23 UT (Figure 1a). The flare exhibited a two-ribbon structure as shown in a snapshot ultraviolet (UV) 1600 \AA\ image (at a pixel size of 0\arcsec.6) taken by the Atmospheric Imaging Assembly (AIA)\cite{Lemen2012} on the Solar Dynamics Observatory (SDO)\cite{Pesnell2012} (Figure 1b). NST/VIS observations in the H$\alpha$ 6563 \AA\ line centre and off-bands ($\pm$0.6~\AA\ and $\pm$1.0~\AA) are also available, each at a time cadence of 28~s. The field-of-view (FOV) of the NST/VIS images is $\sim$57\arcsec$\times$64\arcsec, covering the main portion of the eastern ribbon (Figure 1c and 1d). The pixel size of the NST/VIS images is 0\arcsec.03 ($\sim$20 km), approaching the diffraction limit of the telescope, and 20 times smaller than that of the SDO/AIA UV and EUV images, allowing us to study the fine-scale structures at the chromosphere in unprecedented detail.

Supplementary Movie 1 online shows the NST/VIS H$\alpha$+1.0~\AA\ image sequence (rebinned to 630$\times$706 pixels) running from 17:14 to 19:35 UT at 28~s cadence. In the movie we see the ribbon propagating eastward across the sunspots, the appearance of more and more post-flare loops which are filled with condensing chromospheric plasma, the clumps of chromospheric plasma falling toward the loop footpoints (coronal rain hereafter), and the fine-scale brightenings on the chromosphere when it is impacted by the condensed plasma. We use the original, full resolution images to measure the widths of flare ribbons, post-flare loops and footpoint brightenings, presented in the following.

In the standard flare model, the red-shifted leading edge of the flare maps out the footpoints into which nonthermal electron beams precipitate along the newly reconnected field lines\cite{Svestka1980}. To localize this red-shifted leading edge of the flare, we produced pseudo Dopplergrams in H$\alpha~\pm$1.0~\AA\ by subtracting the red wing from the blue wing images. In these pseudo Dopplergrams, the red shifts (i.e., downward flows) are shown in red color. A red-shifted ribbon can be identified along the outer border of the flare (Figure 2b-2e), and used here to localize the leading edge. As shown in the representative examples in Figure 2, the cross-sectional Gaussian full width at half maximum (FWHM) of the leading edge ranges from 110 to 161 km, very close to the narrowest red-shifted flare ribbon (100 km) detected so far\cite{Sharykin2014}.

While the flare proceeds into a long decay phase, post-flare loops that were filled with hot plasma due to chromospheric evaporation begin to cool and become visible in H$\alpha$. Figure 3a shows a H$\alpha$+1.0~\AA\ snapshot of the post-flare loop system. We selected six distinctive loops (marked by the short cross-cut slits in Figure 3a) in this image and measured their cross-sectional Gaussian FWHM which is within a range of 89--133 km (Figure 3b-3g). The H$\alpha$ post-flare loops found here have nearly the same size as the loops recently detected by another large telescope SST\cite{Antolin2012, Scullion2014}, and represent one of the finest scale loop structures resolved to date. 

The most interesting new phenomenon revealed by this observation is the fine-scale brightenings at the footpoints of falling plasma. It is obvious in the movie that coronal rain streams down along the loops and causes brightenings when it impacts the surface. These brightenings are red-shifted in our pseudo Dopplergrams, also suggesting a connection with the downward flow of plasma (Figure 4c). Here we measure the cross-sectional FWHM of the brightenings, as demonstrated for a few examples in Figure 4. Each brightening usually disappears within 1-2 min with the depletion of the stream of plasma. On the whole they appear in succession with a clear tendency to appear progressively further away from the primary magnetic polarity inversion line (PIL) with time (Figure 5) just like the flaring ribbons do.

Figure 6 presents the FWHM distribution of post-flare loops and that of footpoint brightenings, based on the samples of 107 loops and 108 brightenings. Note that our samples do not cover all the loops and brightenings identified in the image sequence. We tend to select those of distinctive and relatively pronounced cross-sectional intensity profiles close to a Gaussian distribution. As shown in Figure 6, flare loops and footpoint brightenings have similar width distributions with an average of 124$\pm$19 km and 150$\pm$15 km, respectively. 

\section*{Discussion}

The cross-sectional widths of flare ribbons, flare loops and brightenings at the footpoints of coronal rain in this case are mostly in the range of 80-200 km, well below the resolution of most current instruments used for flare studies. Then an important question is whether the loops/brightenings detected by NST/VIS in this study represent a true physical spatial scale of the energy transport mechanism. Unfortunately, even with the most powerful ground-based telescope in the present, we still cannot make such a statement because we cannot exclude the possibility that these loops/brightenings are composed of collections of finer substructures which scale below the resolution limit. In fact, according to some nanoflare models, the cross-sectional width of coronal loop substructures could be as small as 15--20 km\cite{Cargill2004, Vekstein2009}, which is beyond the present observational capabilities.

Nevertheless, getting to such a level of fine structures as shown in this study enables us to move forward in understanding the fundamental spatial scale of flare energy transport mechanism. For instance, using the leading edge of flare ribbon to determine an FWHM of H$\alpha$ emission, we get a width $d$ of 110--161 km (Figure 2). Assuming that the leading edge consists of a single string of circular footpoints, the footpoint area $\pi(\frac{d}{2})^2$ would be of order 10$^{14}$ cm$^2$, which is at least two orders of magnitude smaller than previously reported flare kernel size measured in hard X-ray (HXR)\cite{Dennis2009, Kennedy2015} and at least four times smaller than that in earlier optical observations\cite{Jess2008, Krucker2011, Xu2012a, Xu2016}. This footpoint area is critical in many radiative hydrodynamic models of electron beam precipitation in flares\cite{Abbett1999, Allred2005, Allred2015}. In addition, a multi-threaded chromospheric density model has been proposed to explain the disparity between predicted and observed vertical extents of HXR sources\cite{Kontar2010}. In that model, the chromosphere is not a uniform median but instead is composed of fine-scale strands of different density structure. The fine-scale footpoint area determined by our observation suggests a very small filling factor, hence an extremely high flux of nonthermal electron beams per unit area, which has not yet been dealt with by any radiative hydrodynamic models.

On the other hand, many hydrodynamic loop models also suggest that, in order to form stable and warm coronal loops, coronal loops may consist of unresolved strands, with typical cross-sectional width on the order of 10$-$100 km\cite{Gomez1993, Reale2000, Beveridge2003, Warren2003, Klimchuk2006, Klimchuk2008, Brooks2012}. In other words, most observations of coronal loops may represent superpositions of multiple unresolved strands at different stages of heating and cooling\cite{Klimchuk2006}. A recent study based on analysis of SST/CRISP and SDO/AIA observation confirmed the existence of multi-thermal, multi-stranded substructures, which are mostly on $\sim$100 km spatial scales approaching the diffraction limit of the SST/CRISP \cite{Scullion2014}. Here we detect similarly scaled loop strands with NST/VIS, which is another piece of evidence for the presence of such substructures as was predicted by the models.

We notice that the brightenings at the impact points of coronal rain are different in many respects from other flare-related brightenings, those known as sequential chromospheric brightenings (SCBs)\cite{Balasubramaniam2005, Pevtsov2007, Kirk2012}. SCBs, which usually show no discernible connection with the coronal loops, are presumably associated with a non-localized (sometimes even transequatorial) loop system which is destabilized during the eruption; they appear first near the flare site during the impulsive phase and often precede the H$\alpha$ emission peak; then they appear progressively far from the flare site, with multiple clusters in different directions\cite{Balasubramaniam2005, Pevtsov2007, Kirk2012}. The brightenings found here, in contrast, appear at the footpoints of the post-flare loops simultaneously with the impact of dense plasma, and manifest a nearly organized moving pattern by following the path of a previously observed flaring ribbon. In other words, they behave as a rerun of the flare, but of course at an energetically negligible magnitude. More importantly, the brightenings found here are on a much finer scale than SCBs. Only state-of-the-art large ground-based telescopes such as NST are able to resolve such fine features.

We also notice that small-scale bright footpoints (in and around a sunspot's umbra) of coronal loops were previously detected with Interface Region Imaging Spectrograph (IRIS)\cite{DePontieu2014} in UV channels by Kleint et al.\cite{Kleint2014}. Those footpoints have a spatial FWHM of 0\arcsec.35--0\arcsec.7 (corresponding to $\sim$250--500~km), and exhibit intermittent emission and bursts of strong Doppler red shifts with supersonic velocity up to 200 km s$^{-1}$. There were no flares during that IRIS observation, and Kleint et al. attributed the cause of those bright footpoints to the impact of coronal rain on the transition region (TR). For the event presented here, we also search for such brightenings in IRIS slit-jaw images. However, due to the facts that the pixel scale of IRIS slit-jaw images (0\arcsec.17 pixel$^{-1}$) is about 5 times larger than that of NST/VIS H$\alpha$ images, and these small brightenings were embedded in the remnants of the flare emission, the search is very difficult. Only a few bright dots are identified in the IRIS image sequence in the near UV passband. These brightenings are of FWHM of $\sim$900-1300 km and virtually coinciding in position, time and duration with those in the NST H$\alpha$ images. One example is demonstrated in Figure 7. Unfortunately the IRIS spectrograph slit did not cut though any brightenings. Although without the information on Doppler shifts of spectral lines we cannot fully verify it, it seems reasonable to suppose that the H$\alpha$ brightenings revealed by this observation are chromospheric counterparts to the UV brightenings in the TR reported by Kleint et al.\cite{Kleint2014}. Compared to the previous IRIS observation, our observation captures the finer components in the deeper, cooler and more dense chromosphere during an M-class flare, and clearly demonstrates the cause-and-effect association between falling plasma and brightenings. Moreover, the brightenings following the path of the primary flaring ribbon makes this observation strikingly interesting and unique.

Concerning the physical mechanisms potentially responsible for the brightenings, we generally consider two possible scenarios. In the first scenario, the falling plasma dissipates its kinetic energy into the chromosphere via collision, resulting in the observed brightenings. In this case, the plasma does not decelerate while falling and hits the solar surface usually at several hundred kilometers per second, such as those reported by Gilbert et al.\cite{Gilbert2013} and Reale et al.\cite{Reale2013}. In the second scenario, the thermal non-equilibrium at the loop apex triggers catastrophic cooling of dense plasma as well as causing a pressure imbalance inside loops\cite{Muller2004, Antolin2010}. Driven by pressure difference, siphon-like flows of plasma undergo a transition from super- to subsonic velocities, leading to a shock wave front in the atmosphere\cite{Cargill1980}. We speculate that the second scenario is a more suitable heating mechanism in this case, when considering the preceding flare may be the triggering event of a loss of thermal equilibrium and pressure imbalance. We do see supersonic downflows near the loop footpoints in the TR (figure 7b), but have no quantitative H$\alpha$ Doppler measurement at the chromosphere. Thus we cannot come to a conclusion. In both collision and shock wave scenarios, the local density and temperature are expected to increase, but the energy involved in the process and the resulting brightening area are probably too small to be detected by most current instruments.

To summarize, the high resolution NST observation presented here provides a wealth of detail on the M6.5 solar flare on 2015 June 22. The observation of this quality and completeness of a flare did not exist before. This study focuses on the measurement of the spatial scales of flare ribbons, loops and footpoint brightenings which are essential pieces of information to the full picture of flare energy transport. Our measurements bridge the gap between models and observations as well as opening interesting avenues of future investigation.

\subsubsection*{Acknowledgements}
The data used in this paper were obtained with NST at Big Bear Solar Observatory, which is operated by New Jersey Institute of Technology. BBSO operation is supported by NJIT, US NSF AGS-1250818 and NASA NNX13AG14G, and NST operation is partly supported by the Korea Astronomy and Space Science Institute and Seoul National University, and by strategic priority research program of Chinese Academy of Sciences with Grant No. XDB09000000. This work is supported by NASA under grants NNX11AQ55G, NNX13AG13G and NNX13AF76G, NSF under grants AGS-1153226, AGS-1153424, AGS-1250374, AGS-1348513 and AGS-1408703. We thank the BBSO team for the H$\alpha$ images and thank the NASA/SDO team for the UV filtergrams.

\bigskip

\bigskip

\noindent {\bf \large{Author contributions}}\\
\noindent J. J. carried out the data analysis and wrote the manuscript. Y. X. was the PI of this NST observation and contributed to the scientific ideas. W.C. developed instruments. C.L. contributed to the NST data processing. D. G. contributed to scientific ideas and writing of the paper. H.W. developed the scientific ideas for this study and contributed to the NST data processing. All authors participated in discussions, read and commented on the manuscript.

\bigskip
\bigskip

\subsubsection*{Competing Interests}
The authors declare that they have no competing financial interests.

\subsubsection*{Correspondence} Correspondence and requests for materials should be addressed to J.J. (email: ju.jing@njit.edu)
\clearpage

\begin{figure}
\centering
\includegraphics[scale=0.9]{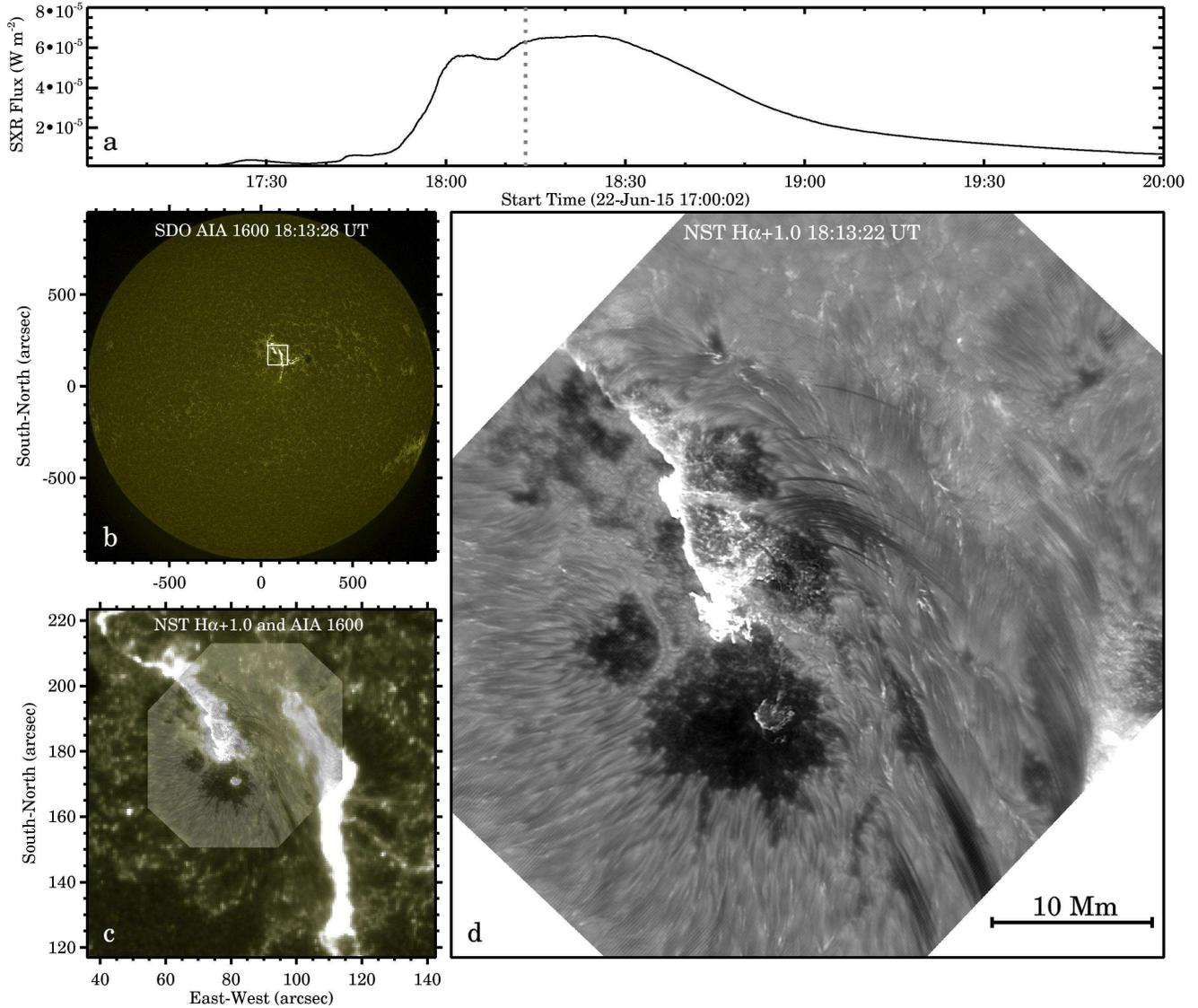}
\caption{Panel a: GOES soft X-ray 1-8 \AA\ light curve. The vertical dotted line indicates the time 18:13:22 UT. The SDO/AIA and NST/VIS images shown in panels b-d were all taken within 6 s from this time. Panel b: the full-disk SDO/AIA 1600 \AA\ map. It serves as the reference to register the NST's FOV with heliographic coordinates. The white box outlines the region of interest (ROI) where the flare occurred. Panel c: a blend of NST/VIS H$\alpha$+1.0~\AA\ image and a larger SDO/AIA 1600 \AA\ map. The field-of-view (FOV) of panel c is the same as the boxed region in panel b. Panel d: the zoomed-in view of the NST/VIS image in panel c. The entire NST/VIS H$\alpha$+1.0\AA\ image sequence can be found as Supplementary Movie 1 online.
\label{f1}}
\end{figure}

\begin{figure}
\centering
\includegraphics[scale=.9]{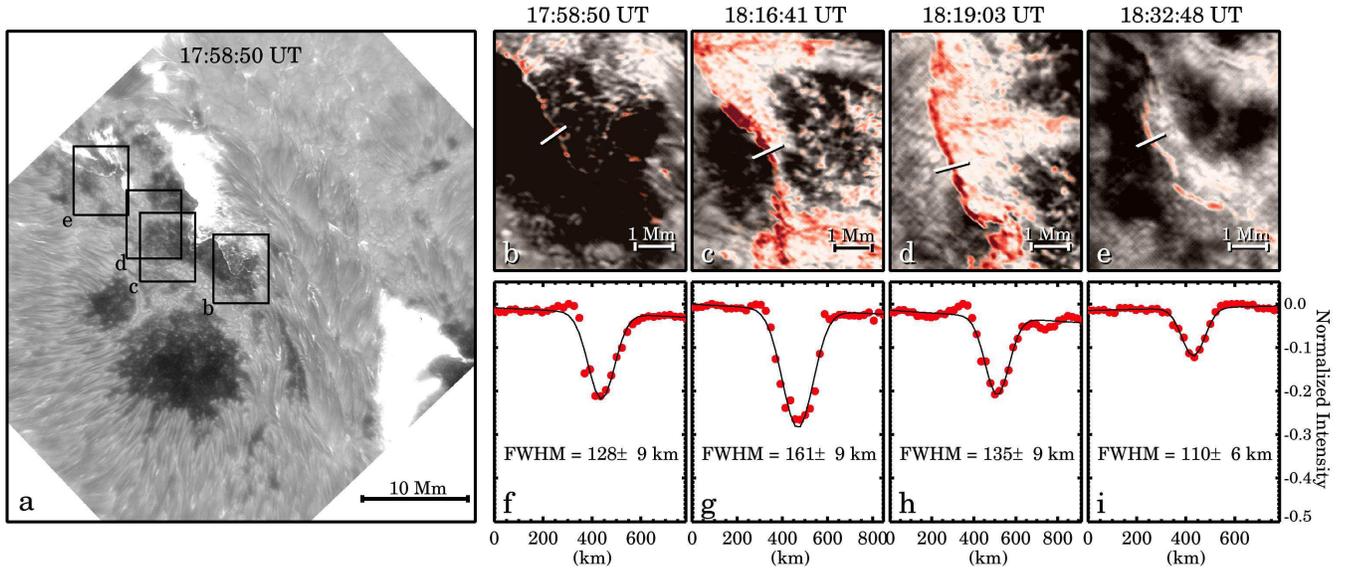}
\caption{Panel a: a snapshot of H$\alpha$+1.0~\AA\ image taken in the impulsive phase of the flare. Four boxes (labelled from `b' to `e') mark the FOV of the zoomed-in images in panels b-e in this figure. Panels b-e: four blends of the H$\alpha$+1.0~\AA\ images (gray scale) and the associated pseudo Dopplergrams (with red corresponding to red shift), at selected times. In each panel the slit cross-cutting the red-shifted leading edge of the flare shows where the cross-sectional width was measured. Panels f-i: the normalized pseudo Dopplergram intensity profile (red dots) along the slit in the top panel and the Gaussian fit (black curve). The Gaussian FWHM and $\pm3\sigma$ are provided in each panel.
\label{f2}}
\end{figure}

\begin{figure}
\centering
\includegraphics[scale=.9]{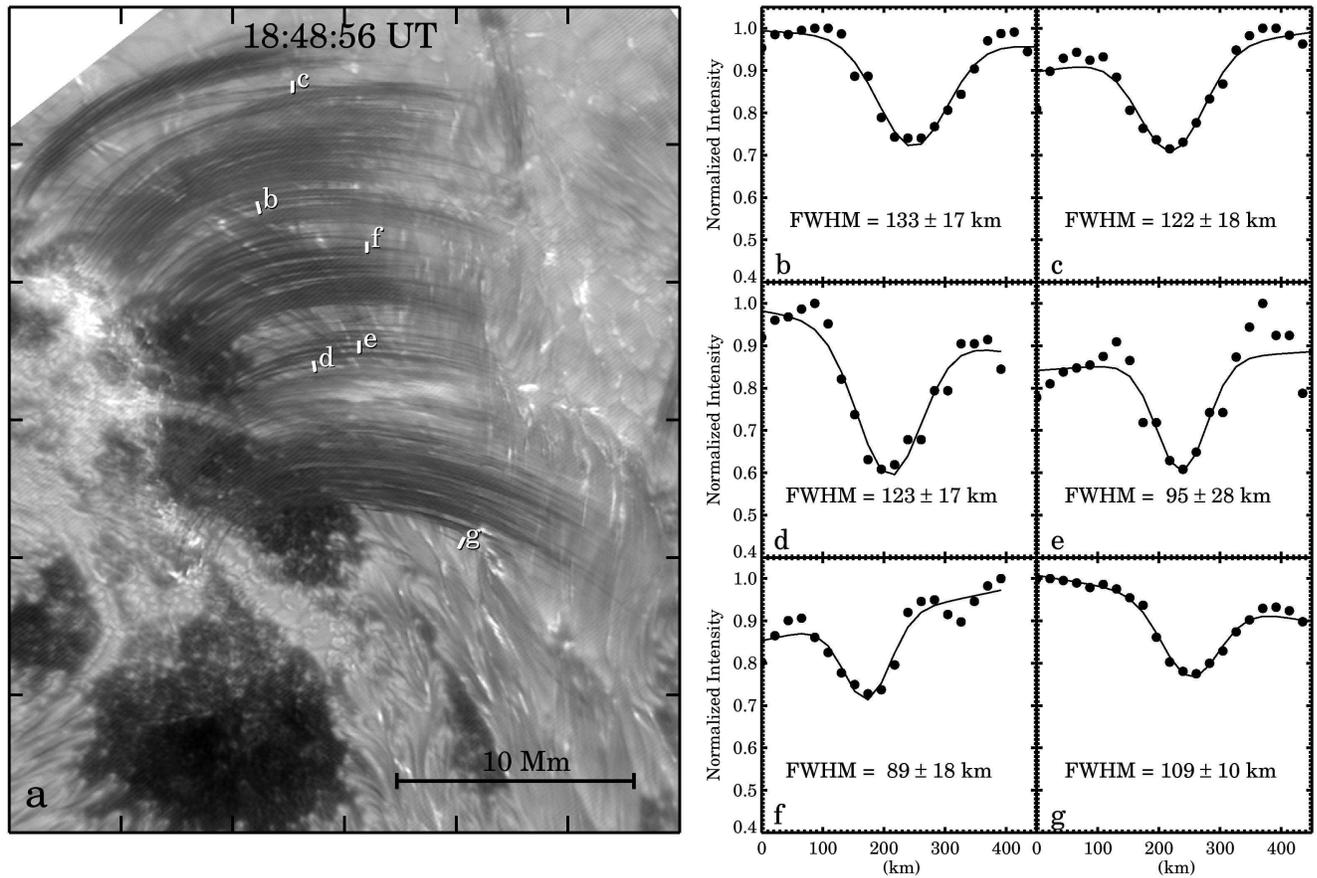}
\caption{Panel a: a snapshot of H$\alpha$+1.0~\AA\ image taken in the decay phase of the flare. Six short slits mark where the cross-section of six loops were measured. Panels b-g: the normalized H$\alpha$+1.0~\AA\ intensity profiles (black dots) along the slits and the Gaussian fits (black curves). The six slits in panel a are distinguished by labelling with panel letters `b' through `g'. The Gaussian FWHM and its $\pm3\sigma$ are provided in each panel.
\label{f3}}
\end{figure}

\begin{figure}
\centering
\includegraphics[scale=.9]{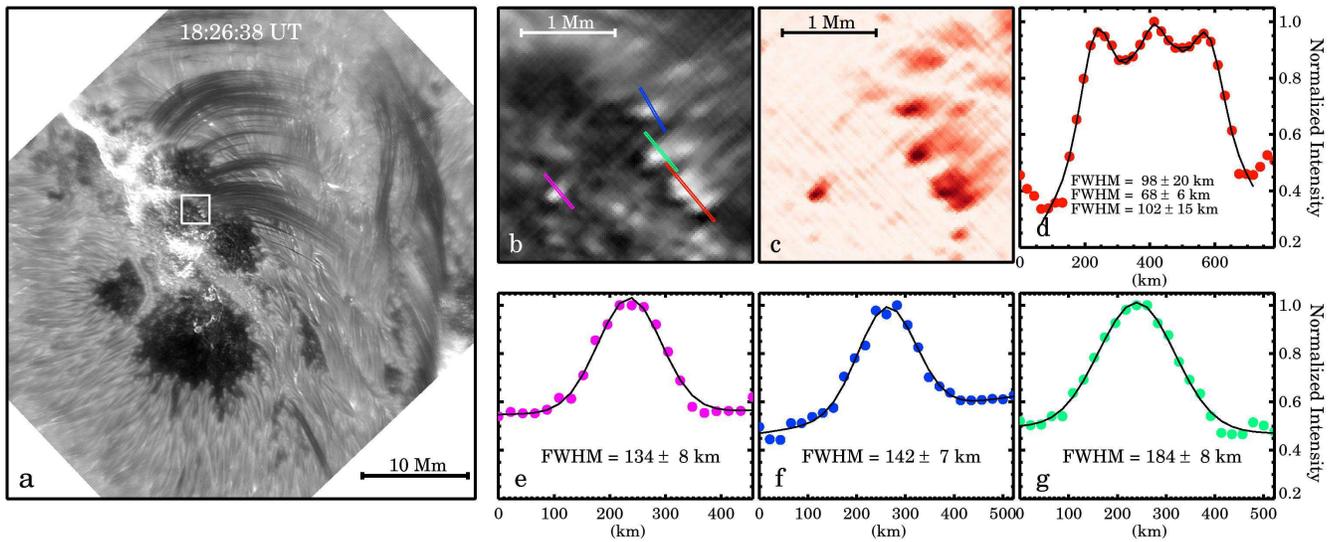}
\caption{Panel a: a snapshot of H$\alpha$+1.0~\AA\ image showing post-flare brightening associated with coronal rain. The white box marks the ROI where the brightenings occurred. Panel b: the zoomed-in view of the ROI marked by the white box in panel a. The slits, with different colors, mark where the brightenings' cross-sectional width were measured. Panel c: the pseudo Dopplergrams (with red corresponding to red shift) of the ROI. Panels d-g: the normalized H$\alpha$+1.0~\AA\ intensity profiles along the slits and the Gaussian fits. For reference, the color of each profile is the same as the color of the slit (see panel b). The Gaussian FWHM and $\pm3\sigma$ are provided in the panels.
\label{f4}}
\end{figure}

\begin{figure}
\centering
\includegraphics[scale=.85]{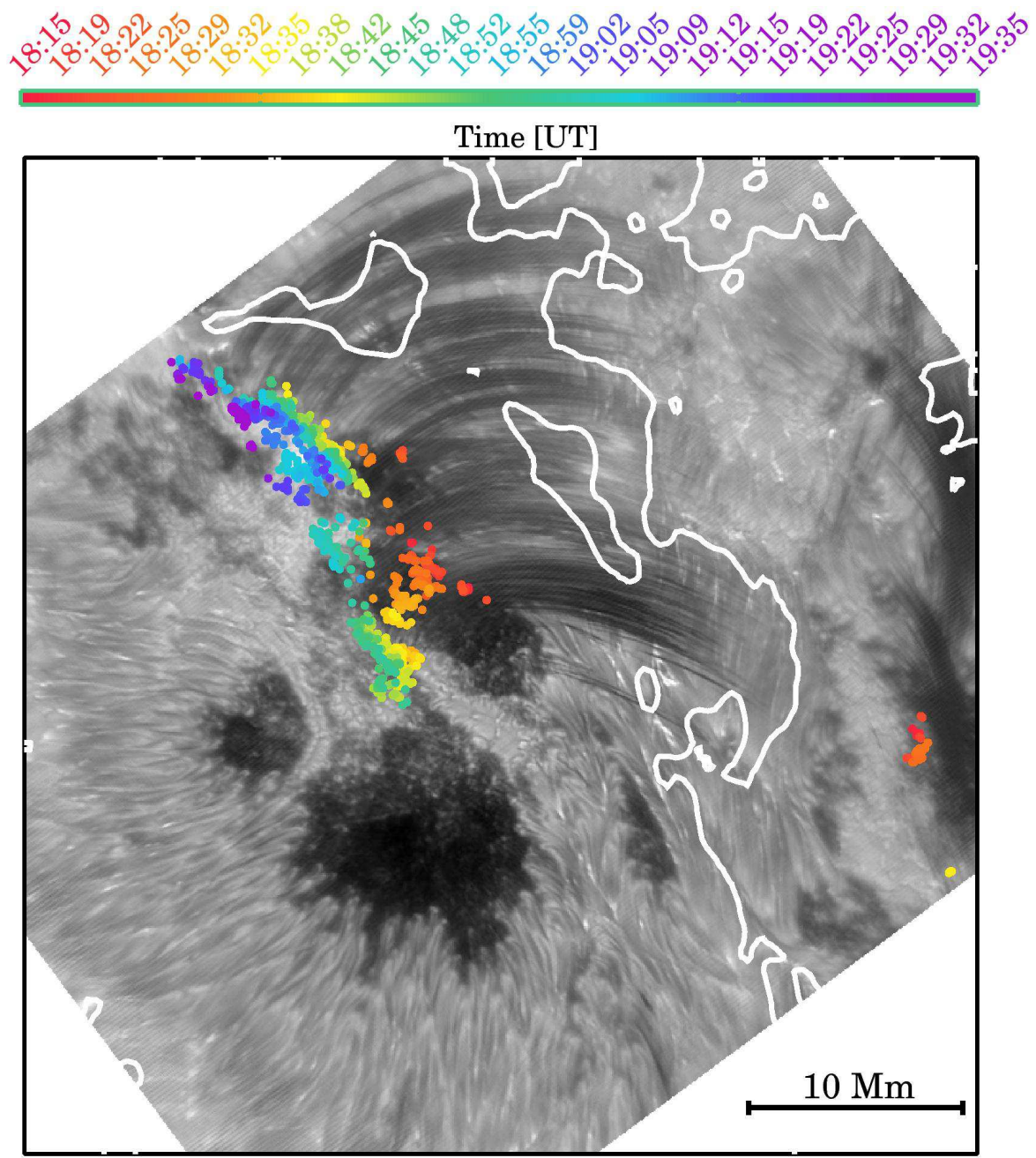}
\caption{A snapshot NST H$\alpha$+1.0~\AA\ image (gray scale) taken at 18:54:34 UT. White contours are magnetic polarity inversion lines (PILs). The superimposed dots summarize the locations of footpoint brightenings in the image sequence which were identified manually. The colors are assigned in chronological order of appearance. Time is indicated by the color code.
\label{f5}}
\end{figure}

\begin{figure}
\centering
\includegraphics[scale=.6]{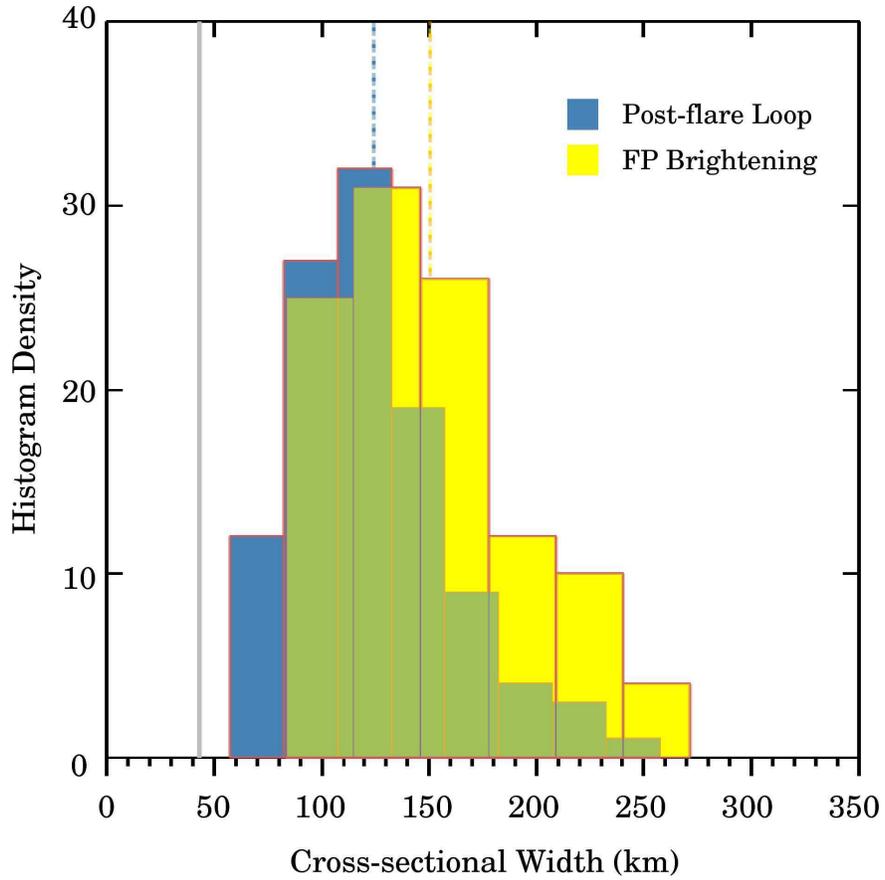}
\caption{Cross-sectional Gaussian FWHM distribution of post-flare loops (blue) based on a sample of 107 loops, and that of footpoint(FP) brightenings (yellow) based on a sample of 108 brightenings. The vertical blue/yellow dashed lines represent the mean of each distribution. The vertical grey solid line marks the resolution limit of NST/VIS.
\label{f5}}
\end{figure}

\begin{figure}
\centering
\includegraphics[scale=.85]{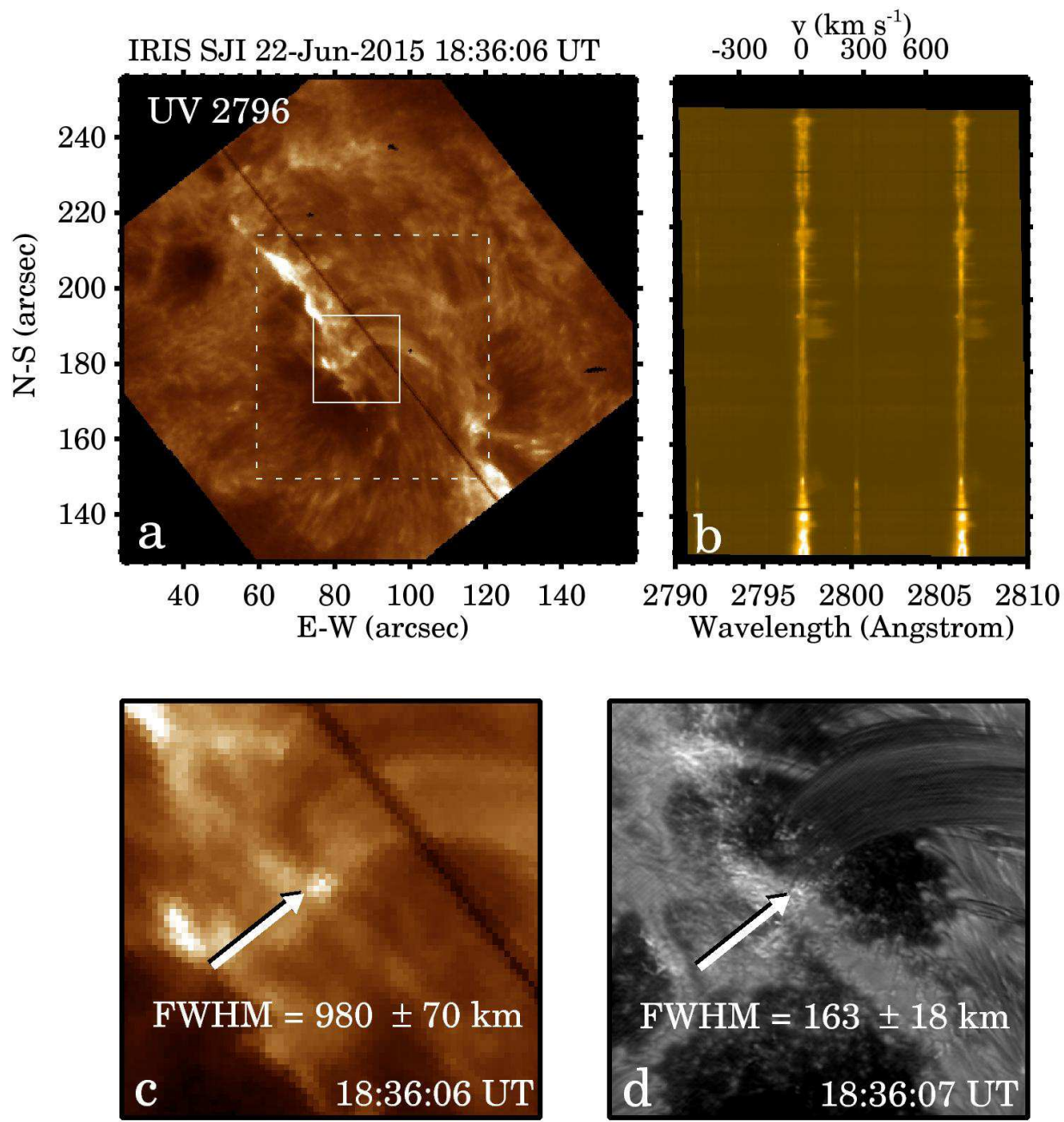}
\caption{Panel a: an IRIS slit-jaw image obtained with the 2796~\AA\ (Mg~{\sc ii} k) filter. The black line is the spectrograph slit. The dashed box marks the FOV of NST H$\alpha$ images, and the solid box marks the FOV of panels c and d. Panel b: Doppler shifts and velocity of Mg~{\sc ii} k spectral line. Panels c--d: the zoomed-in IRIS Mg~{\sc ii} k slit-jaw image and the NST H$\alpha$ image, both have the same FOV and were taken almost at the same time. The brightenings are indicated with arrows. The Gaussian FWHM and $\pm3\sigma$ are provided.
\label{f5}}
\end{figure}

\end{document}